\documentclass[11pt]{article}
\usepackage{jheppub}

\usepackage{slashed}
\usepackage{graphicx}
\usepackage{subfigure}
\usepackage{dcolumn}
\usepackage{bm}

\usepackage{amsfonts}
\usepackage{xcolor}

\newcommand{\newc}{\newcommand}
\newc{\eeq}{\end{equation}}
\newc{\beq}{\begin{equation}}
\newc{\eeqa}{\end{eqnarray}}
\newc{\beqa}{\begin{eqnarray}}
\newc{\nonr}{\nonumber}
\newc{\bi}{\begin{itemize}}
\newc{\ei}{\end{itemize}}
\newc{\ra}{\rightarrow}
\newc{\st}{s_\theta}
\newc{\ct}{c_\theta}
\newc{\SL}{\not\!\!}

\mathchardef\mhyphen="2D

\begin{document}

\title{Prospects for Detecting light bosons at the FCC-ee and CEPC}
\author[a,b]{We-Fu Chang}

\author[b]{John N. Ng}

\author[b]{Graham White}%
\affiliation[a]{ Department of Physics, National Tsing Hua University, Hsinchu 300, Taiwan}
\affiliation[b]{TRIUMF Theory Group, 4004 Wesbrook Mall, Vancouver, B.C. V6T2A3, Canada}
\emailAdd{wfchang@phys.nthu.edu.tw}
\emailAdd{misery@triumf.ca}
\emailAdd{gwhite@triumf.ca}
\date{\today}

\abstract{
We look at the prospects for detecting light bosons, $X$, at proposed Z factories assuming a production of $10^{12}$ Z bosons. Such a large yield is within the design goals of future FCC-ee and CEPC colliders.  Specifically we look at the cases where $X$ is either a singlet scalar which mixes with the standard model Higgs or a vector boson with mass $1\lesssim M_X \lesssim 80$ GeV.   We find that several channels are particularly promising for discovery prospects. In particular  $Z\rightarrow f \bar{f} X$ and $Z \rightarrow V_Q X$ gives a promising signal above a very clean standard model background. We also discuss several channels that have too large a background to be useful.
}

\maketitle

\section{Introduction}
So far the LHC has not uncovered any unambiguous evidence for physics beyond the standard model(SM). It is striking then to consider that in spite of this impressive advance at the energy frontier how relatively unconstrained the parameter space is for new bosons that are lighter than a Higgs.  This will remain true even in the case where the high luminosity LHC fails to find new physics \cite{Chang:2017ynj}. Two proposed experiments could either result in detection or falsify large parts of the parameter space for such a light boson, the FCC-ee and the CEPC\cite{CEPC-SPPCStudyGroup:2015csa,CEPC-SPPCStudyGroup:2015esa,Gomez-Ceballos:2013zzn,dEnterria:2016sca,dEnterria:2016fpc}. It is within the design goals of both experiments to produce up to $10^{12}$ Z bosons per year making the even rare decays probable. Such rare decays have been proposed as a way to constrain a hidden dark sector \cite{Liu:2017zdh,Wu:2017kgr}, as an indirect probe for supersymmetry \cite{Wu:2017kgr} and a probe to an electroweak phase transition \cite{Huang:2016cjm}.

In this work we consider the singlet extension of the standard model as well as a new vector boson which  couples to the standard model through effective operators.
Such new bosons are ubiquitous in extensions to the standard model  \cite{Chang:2017ynj,Chang:2012ta,Ng:2015eia,Ellwanger:2010oug,Athron:2009bs,Balazs:2013cia,Akula:2017yfr,Ivanov:2017dad,Fink:2018mcz,Vieu:2018nfq,Croon:2015naa,Aydemir:2015nfa,Aydemir:2016xtj,Croon:2015wba}. New scalar particles can be a dark matter candidate, a portal to the dark sector \cite{Chang:2016pya,Chang:2014lxa,Cabral-Rosetti:2017mai,Garg:2017iva,McKay:2017iur,Athron:2017kgt,Burgess:2000yq,McDonald:1993ex,Cline:2013gha} as well as a catalyzing a strongly first order electroweak phase transition \cite{Profumo:2014opa,Profumo:2007wc,White:2016nbo,Kozaczuk:2014kva} and improving the stability of the vacuum \cite{EliasMiro:2012ay,Gonderinger:2009jp,Gonderinger:2012rd,Khan:2014kba,Balazs:2016tbi}.  We find that one can discover such a light scalar even for relatively small mixing angles of $\sin^2 \theta \sim 10^{-7} \to 10 ^{-3}$ with the SM Higgs boson depending on the mass and  the decay branching ratio $ Br(S\ra\mbox{final state})$. This impressive search power is due to relatively clean
SM backgrounds for decay channels $Z \to X f \bar{f} $ and $Z \to V_Q X$ where $V_Q$ is the $1^3S_1$ quarkonium with $Q \in (c,b)$. In both of these cases the SM backgrounds are dominated by $X$ decaying into either $b \bar{b}$, $\mu \bar{\mu}$ or invisible final states (neutrinos, or massless Goldstone bosons or yet unknown dark matter). Such channels turn out to the most promising for reasons we discuss throughout the paper. We perform a systematic analysis of every possible decay channel of these types which we list in table \ref{tab:SMBG}. In the case of spin-1 boson extensions to the standard model the limits we find depends on which fermion it couples to as well as its mass.  The Z factories have the potential to directly detect a few GeV vector boson whose gauge coupling strength to the SM fermion is as small as $\sim 10^{-4} e$. 
 Throughout this paper, both numerical and analytic methods are used with the numerical study relying heavily on the numerical package CalcHep\cite{Belyaev:2012qa}. 
 
The structure of this paper is as follows. We discuss the singlet and spin-1 extensions of the standard model in section \ref{sec:models}, then discuss the two most promising decay channels including an analysis of excesses over the SM expectations  as a function of the parameter space and the standard model background in sections \ref{sec:1stdecay} and \ref{sec:quarkoniumdecay} respectively. Next we discuss the case where the singlet does not mix in \ref{sec:vevless}. Finally we compare the signal to the background in section \ref{sec:conclusions}. We end with a conclusion.

\section{Scalar and spin-1 boson extensions of the standard model}\label{sec:models}

Let us begin with a scalar singlet ($S$) extension to the standard model. The singlet couples to the standard model through a Higgs portal term. A commonly used one is  $S^\dagger S H^\dagger H$ where  $H$ is the SM Higgs field. As such the decay rates of $Z\ra S f\bar{f}$ or $Z\ra V_Q S$ only depend on the physical mass, $M_S$, of the singlet scalar and its mixing angle with the SM Higgs.
 $S$ is produced on shell and decays either visibly into SM particles or into some invisible final states.
One will then observe a resonant peak in the invariant mass of the final states. Note that the invariant mass of the invisible decay can be well determined in the rest frame of Z boson.
The specific signal, $S\ra Y$, is then controlled by the model dependent branching ratio the branching ratio $Br(S\ra Y)$.
We aim to calculate prospective limits in the $\sin^2\theta \times Br(S\ra Y)$ v.s $M_S$ plane at future Z factories.
Note that these limits are independent of the details of the model.

To set up our conventions, we focus only on the relevant Lagrangian for a scalar field extension that is a singlet under the standard model gauge group and the interaction with the standard model is through the general Higgs portal.
We denote the weak basis of the real components of SM Higgs and the singlet as $(h_0, s_0)$. For our purposes we will only need the mass squared matrix
\beq
\left(
  \begin{array}{cc}
  m^2_{h_0} & m^2_{sh}\\ m^2_{sh} & m^2_{s_0}
  \end{array}
\right)\,.
\eeq
We stress that the origins of the mass squared matrix is irrelevant to this work and it can arise from either explicit or spontaneous symmetry breaking.
The mass matrix can be diagonalized by a rotation
\begin{eqnarray}
\left(\begin{array}{c}  H_m\\ S_m \end{array} \right)
 = \left( \begin{array}{cc} \cos \theta &-\sin \theta \\ \sin \theta & \cos \theta
      \end{array} \right)
      \left(\begin{array}{c} h_0\\ s_0  \end{array}  \right)
\end{eqnarray}
with mixing angle $\theta$, and
 \beq
 \tan 2\theta ={2 m^2_{sh}\over  m^2_{s_0} -m^2_{h_0} }\,.
 \eeq
Here the range of the mixing angle is $\theta\in[-\pi/2, \pi/2]$.
 We consider the case where $ m^2_{s_0}< m^2_{h_0}$
and the mass eigenvalues are
 \beqa
M_{H_m}^2=\frac{1}{2}\left[  m^2_{h_0} +  m^2_{s_0} +\sqrt{(m^2_{h_0}- m^2_{s_0})^2+ m^2_{sh}}\, \right] \,,\nonr\\
M_{S_m}^2=\frac{1}{2}\left[  m^2_{h_0}+ m^2_{s_0}-\sqrt{(m^2_{h_0}- m^2_{s_0})^2+ m^2_{sh}}\, \right]\,.
 \eeqa
In the absence of mixing, i.e. when $ m^2_{sh}=0$, one has $ M_{H_m}= m_{h_0}$ and $ M_{S_m}= m_{s_0}$.

$H_m$ is  identified as the $M_H=125$ GeV SM-like Higgs  boson and $S_m$ is the new neutral scalar boson with unknown mass $M_S$.   For notational simplicity we shall drop the subscript $m$ for the mass eigenstates and use the shorthand notations $s_\theta, c_ \theta$ for
$\sin \theta, \cos \theta$. We are interested in the mass range $ 1\lesssim M_S\lesssim 80$ GeV. The upper bound is limited by kinematics and the lower bound of $M_S\simeq 1$ GeV is due to  the energy resolution of the future Z factories which we assume to be around $1$GeV.

Finally we also consider the case of a new spin-1 boson with the same mass range.
For simplicity, we only consider the case where the interaction Lagrangian between the spin-1 boson and the standard model  is phenomenologically parameterized  as
\begin{equation}
{\cal L} \supset (e g_D^f)\bar{f}\gamma_\mu f V_D^\mu
     \ .
\end{equation}
The UV completion of such a model is not our concern in this work. See refs. \cite{Chang:2006fp,Chang:2007ki} for examples of plausible UV completions that give rise to the above operator. \par
In this case we have only two free parameters in the mass of the new boson and the coupling strength $g_D^f$  for each flavor.   For the cases with axial vector couplings, the study can be easily extended with one more free parameter for each flavor.
In general, the new vector boson could acquire flavor changing couplings.  We will only focus on the case that the new vector boson couplings are  flavor conserving.
However, our study can be trivially extended to the flavor changing case where the SM background can be ignored and some useful new limits can be placed.

\begin{table}[ht!]
\centering
\renewcommand{\arraystretch}{1.45}
\begin{tabular}{|c|c|c|}
  \hline
  $Z$ decay &  subsequent $X$ decay  & SM background  \\
  \hline
  $Z\to \nu_i\bar{\nu_i} X $ & $X\ra b\bar{b}$ & $Z\ra \underline{b\bar{b}}\ , \nu_i\bar{\nu_i} $ \\
                             & $X\ra \mu\bar{\mu}$ & $Z\ra \underline{\mu\bar{\mu}}\ , \nu_i\bar{\nu_i} $ \\
 \hline
  $Z\to b\bar{b} X $ & $X\ra b\bar{b}$ & $Z\ra \underline{b\bar{b}}\ ,  b\bar{b} $ \\
                     & $X\ra \mu\bar{\mu}$ & $Z\ra \underline{\mu\bar{\mu}}\ ,  b\bar{b} $ \\
                     & $X\ra$ invisible  & $Z\ra \underline{\nu_i\bar{\nu_i}}\ ,  b\bar{b} $ \\
 \hline
  $Z\to \mu\bar{\mu} X $ & $X\ra b\bar{b}$ & $Z\ra \underline{b\bar{b}}\ ,  \mu\bar{\mu} $ \\
                     & $X\ra \mu\bar{\mu}$ & $Z\ra \underline{\mu\bar{\mu}}\ ,  \mu\bar{\mu} $ \\
                     & $X\ra$ invisible  & $Z\ra \underline{\nu_i\bar{\nu_i}}\ ,  \mu\bar{\mu} $ \\
  \hline
  $Z\to J/\psi  X $  & $X\ra \mu\bar{\mu}$ & $Z\ra J/\psi \underline{\mu\bar{\mu}} $ \\
                     & $X\ra$ invisible  & $Z\ra J/\psi \underline{\nu_i\bar{\nu_i}} $ \\
  \hline
   $Z\to \Upsilon  X $  & $X\ra \mu\bar{\mu}$ & $Z\ra \Upsilon\underline{\mu\bar{\mu}} $ \\
                     & $X\ra$ invisible  & $Z\ra \Upsilon \underline{\nu_i\bar{\nu_i}} $ \\
  \hline
\end{tabular}
  \caption{Possible signals and their SM background. Fake signals within the SM are those whose invariant mass of underlined fermion pair is close to $m_X$.}
\label{tab:SMBG}
\end{table}
\section{Decay channel $Z\to X f\bar{f}$}\label{sec:1stdecay}

In this section we discuss the first promising decay channel where $X$ can be an invisible SM background ($\nu \bar{\nu })$, a visible SM background such as $f \bar{f}$ or the singlet particle we are searching for. For the case of fermions we will only consider bottom quarks and muons. The former due to b tagging and the latter due to its detectability arising from its long lifetime. Light jets and tau leptons turn out to be too noisy to compete with these channels.
We will systematically study the decay rate as a function of the three parameters: the mass and the mixing angle( gauge coupling) of the light scalar( vector) as well as the decay branching fraction of $X$. In the case of the mixing angle(gauge coupling) the signal branching ratios are trivially proportional to $\sin ^2 \theta ( (g_D^f)^2 ) $ and we can simply divide by this quantity to give two parameter plots.
The $Z$ can be produced almost at rest by precisely controlling the energies of $e^\pm$ beams.
The invariant mass of $X$ can be determined by the squared of the 4-momentum sum of all its decay products. We are looking for a resonant invariant mass which peaks at $M_X$.
On the other hand, even if $X$ decays invisibly, for example in the process $Z(P_Z)\to f(k_1) \bar{f}(k_2) X(P_X)$, the invariant mass of $X$
can still be  reconstructed by kinematics of the fermion pair in the $Z$ rest frame.
\begin{eqnarray}
P_X^2&=& [P_Z -(k_1+k_2)]^2 \nonumber \\
&=& M_Z^2+ M_{ff}^2 -2M_Z(E^{CM}_1+E^{CM}_2)\ .
\end{eqnarray}
The SM $Z\to f\bar{f}f'\bar{f'}$ background are calculated by CalcHep and summarized in Fig.\ref{fig:SMBG_Z4f}.

  In $e^+e^-$ collisions at a resonance such as at the Z-pole
 there is always an intrinsic non-resonance contribution to the final states.
Most notedly, when the final states involve charged fermions pair they can originate from virtual photon emissions from the initial states and t-channel processes with a very forward light fermion pair.
 We use CalcHep to evaluate the full SM tree-level cross section $\sigma(e^+e^-\ra \mbox{ final states})$ and we scan the c.m. energy
within $M_Z\pm 10$GeV to extract the continuous non-resonance SM background.

We found that the non-resonance background is indeed much smaller than the
Z-resonance background. For example, the non-resonance to resonance ratios in $e^+e^-\ra \mu \bar{\mu} b\bar{b}$ are
$\{6.5,6.1,13.5\}\%$ for $b\bar{b}$ invariant mass $m_{bb}=\{70\pm1, 45\pm 1, 20\pm1 \}$ GeV, and
$\{4.3, 8.3, 10.5, 17.0\}\%$ for muon pair $m_{\mu\mu}=\{70\pm1, 45\pm 1, 20\pm1, 10\pm1 \}$ GeV.
It is clear that the non-resonance contribution is larger for smaller $m_{\mu\mu}$ where virtual photon mediation is the main process. We obtained $\gtrsim 20\%$ corrections from non-resonant background for $m_{\mu\mu}<10$ GeV only.
For cases with invisible final states such as   $e^+e^-\ra \nu \bar{\nu} b\bar{b}$
with  $\{4.3,2.8,7.2\}\%$ for $b\bar{b}$ invariant mass $m_{bb}=\{70\pm1, 45\pm 1, 20\pm1 \}$ GeV,
and $\{3.5, 2.8, 3.5\}\%$ for neutrino pair $m_{\nu\nu}=\{ 45\pm 1, 20\pm1, 10\pm1 \}$ GeV.
Since the neutrino pair can only come from virtual $Z$ the non-resonance background is consistently small.
Finally, we remark that the signal $X f\bar{f}$ can also have non-resonance contributions. But, these are typically much smaller than the contributions we calculated.

\subsection{$X=S$}
\begin{figure}
    \centering
    \includegraphics[width=0.54\textwidth]{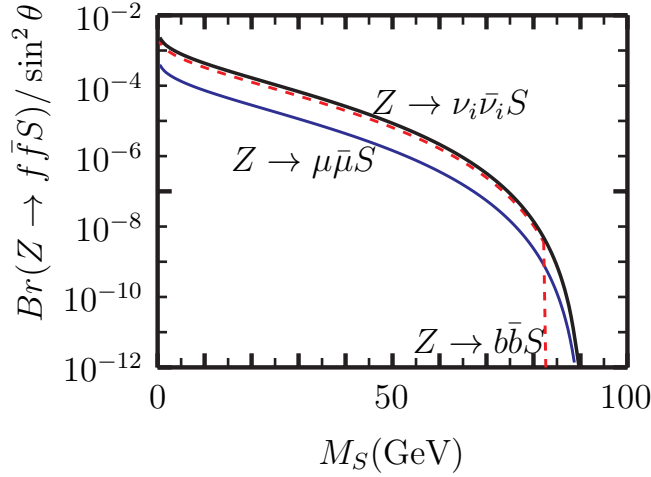}
    \caption{The decay branching ratios of $Z\ra f\bar{f} S$ over the mixing angle squared vs $M_S$.
    The red solid/blue dash/black line is for $f=\mu /b/\nu_i$. }
    \label{fig:Z2Sff}
\end{figure}

Let us begin with the case where $X$ is the light scalar we are ultimately searching for.
Due to the Yukawa suppression, we can ignore diagrams with $S$ attached to the fermion line.
If $M_S>2m_b$, the dominant decay channel is $S\ra 2b$.
A useful kinetic variable $y_b\equiv m^2_{bb}/M_Z^2$  is defined where $m_{bb}$ is the invariant mass of the $b\bar{b}$ pair.
The on-shell light scalar gives a very narrow resonance peak in $y_b$ at around $y_b=(M_S/M_Z)^2$ and stands out from the continuous SM background.
In this case we can calculate the branching ratio analytically and find it to be
\begin{equation}
Br(Z\ra S f\bar{f})= \st^2 \times F(M_S/M_Z) \times Br(Z\ra f\bar{f})\  ,
\end{equation}
where
\begin{eqnarray}
F(x)=\frac{G_F M_Z^2 }{24\sqrt{2} \pi^2 }
&& \left[ {3 x (x^4-8x^2+20)\over \sqrt{4-x^2}}\cos^{-1}\left( \frac{x}{2}(3-x^2)\right) - \right.\nonumber \\
&& \left. 3(x^4-6x^2+4)\ln x-\frac{1}{2}(1-x^2)(2x^4-13x^2+47) \right]\ ,
\end{eqnarray}
Using the PDG\cite{Patrignani:2016xqp} we can acquire the relevant standard model branching ratios \begin{eqnarray}
&& Br(Z\ra b\bar{b})=15.12\%\,, \nonumber \\
&& Br(Z\ra \nu_i\bar{\nu_i})=20.0\%\,, \nonumber \\
&& Br(Z\ra \mu\bar{\mu})=3.366\% \ .
\end{eqnarray} The prediction for the $Z\to f\bar{f}S$ branching ratios normalized by $\sin ^2 \theta $ we show in  Fig.\ref{fig:Z2Sff}.
In the case where the singlet mass is less than twice of the bottom mass, $M_S<2 m_b$, the cleanest signal will be $S \to ee, \mu\mu$ which can be used to reconstruct the $S$ resonance using the quantities $y_e$ and  $y_\mu$. \par
For the case of $S$  decaying into invisible final states, the SM background is $Z\to \nu\bar{\nu} f\bar{f}$. As discussed before, the invariant mass of $X$ can still be reconstructed from the other visible fermions since the $Z$ boson can be produced nearly at rest.

\subsection{$X=V_D$}
The decay width of $V_D\to ff$ can be derived to be
\begin{equation}
\Gamma _{V_D\ra ff}={N_c^f \alpha (g^f_D)^2\over 3}M_{V_D} (1+2 r_f^2)\sqrt{1-4 r_f^2}\ ,
\end{equation}
where $r_f= m_f/M_{V_D}$. Note that there is no tree-level $V_D Z Z$ coupling and the form factor has to be symmetrized for the two $Z$'s.
 By labeling the process as $Z(q,\mu)\to f(p_1) \bar{f}(p_2) V_D(k,\nu)$, where the $V_D$ is attached to the fermion line, the amplitude reads
 \begin{eqnarray}
i{\cal M}=-i \bar{u}(p_1)\left[{ -\SL k\gamma^\nu \gamma^\mu +2p_1^\nu \gamma^\mu \over m_{V_D}^2 +2 p_1\cdot k}-{ -\gamma^\mu \gamma^\nu\SL k +2p_2^\mu \gamma^\nu \over m_{V_D}^2 +2 p_2\cdot k}\right]\nonumber \\
 \times \left(\frac{e g_2 g_D^f}{c_W}\right)(g^f_V-g^f_A \gamma^5 ) v(p_2)\epsilon_\nu^*(k) \epsilon_\mu(q)\ ,
 \end{eqnarray}
after applying the equation of motions and $(\epsilon\cdot k)=0$.
Here $g^f_A=T^f_3/2$ and $g_V=T^f_3/2-q^f s_W^2$ are the SM Z couplings.
The complete analytic expression for cross section is too complicated for practical use.
For $f\neq b$, the $m_f=0$ limit is a good approximation, and the simplified differential decay width reads:
\begin{eqnarray}
\frac{d^2 \Gamma _{Z \to V_D  f\bar{f}}}{dx dy} &=& {\alpha (g_D^f)^2 G_F M_Z^3\over 3\sqrt{2} \pi^2}{N_c^f [(g^f_V)^2+(g^f_A)^2 ]\over y^2(1+d-x-y)^2}\nonumber \\
&& \times\left[ 4y^3(1+d-x) +y(1+d-x)(1+x^2+d^2+4d)-2y^4\right.\nonumber \\
&& \left.-d(1+d-x)^2-y^2(3+3d^2+3x^2+8d-4x-4xd) \right]
\end{eqnarray}
where $d=(M_{V_D}/M_Z)^2$, $x\equiv (p_1+p_2)^2/M_Z^2$ and $y\equiv (p_2+k)^2/M_Z^2$.
The kinematics are $ 0\leq x\leq (1-\sqrt{d})^2$ and  $ (1+d-x)/2 -\lambda_{cm}(x,d)/2\leq  y\leq (1+d-x)/2 +\lambda_{cm}(x,d)/2$ where
 $\lambda_{cm}(x_1,x_2)\equiv \sqrt{1+x_1^2+x_2^2-2(x_1+x_2+x_1 x_2 )}$.
The y-integration can be performed analytically, but we do not find a closed form expression for the double integration.
Instead,  we will just evaluate the complete differential decay rate, with $m_f\neq 0$, numerically.

\begin{figure}
    \centering
    \includegraphics[width=0.54\textwidth]{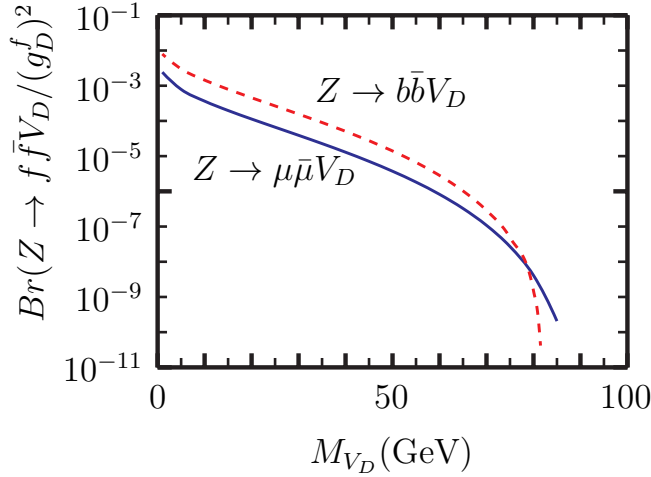}
    \caption{The decay branching ratio of $Z\ra ff V_D$ over the coupling strength squared  v.s. $M_{V_D}$. }
   \label{fig:Z_VDff}
\end{figure}

The widths of $Z\ra V_D f\bar{f}$ are displayed in  Fig.\ref{fig:Z_VDff}

\section{Decay channel $Z\to V_{Q} X$}\label{sec:quarkoniumdecay}
\begin{figure}
    \centering
    \includegraphics[width=0.44\textwidth]{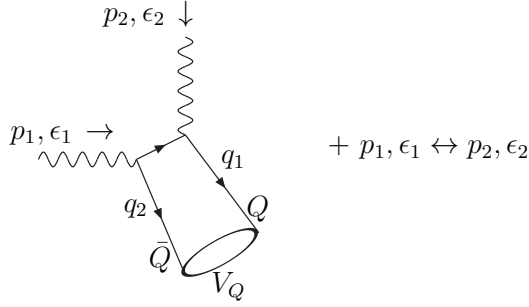}
    \caption{The Feynman diagrams for the $ZZV_Q$ coupling. }
   \label{fig:ZZV_vertex}
\end{figure}

Before considering the process $Z\to V_Q X$, we shall study the SM background $Z\ra V_Q f\bar{f}$, $(f=\mu,\nu)$.
For simplicity, we will not consider $f=c, b$ which are complicated by QCD.
We need to derive the $ZZV_Q$ coupling vertex for calculating the SM process $Z\to V_Q \nu_i\bar{\nu}_i$.  Another SM background $Z\ra \nu\bar{\nu} Z^*(Z^*\ra V_Q)$ is negligible.
Following \cite{Guberina:1980dc},  the desired vertex can be derived from the Feynman diagrams, Fig.\ref{fig:ZZV_vertex}.
Labeling the momenta and polarizations as $Z(p_1,\epsilon_1)-Z(p_2,\epsilon_2)-V_Q(P_V,\epsilon_V)$, $(p_1+p_2=p_V)$,
the reduced amplitude (without the quark spinor wave functions ) reads
\begin{eqnarray}
i{\cal \widetilde{M}}&=& i\frac{g_2^2/c_W^2}{(p_1\!\cdot\! p_2)}\left[\SL \epsilon_2 (g^Q_V\!-\!g^Q_A\gamma^5)(\SL q_1-\SL p_2+m_Q)\SL \epsilon_1(g^Q_V\!-\!g^Q_A\gamma^5)\right.\nonumber \\
&& \left.+ \SL \epsilon_1 (g^Q_V\!-\!g^Q_A\gamma^5)(\SL p_2-\SL q_2+m_Q)\SL \epsilon_2(g^Q_V\!-\!g^Q_A\gamma^5)\right]\  ,
\end{eqnarray}
where $g^Q_V$ and $g^Q_A$ are the SM heavy quark-Z couplings.
In the NR limit, $q_1\sim q_2 \sim p\equiv p_V/2$ and $q_1^2\sim q_2^2\sim m_Q^2$.
Including the spin projection and the quarkonium wave function, the coupling is
\begin{eqnarray}
i{\cal M}& \simeq &-{\sqrt{N_c m_V}\phi_0^Q\over 2 m_Q} Tr[ i{\cal \widetilde{M}}\SL \epsilon_V^*(\SL p+m_Q)]\\
&=& { A^Q_{ZZV}\over ( p_1\!\cdot\! p_2)} \epsilon^{ \epsilon_V, \epsilon_1, \epsilon_2,  p_1-p_2}\ ,
\end{eqnarray}
where  $\phi_0^Q$ is the wave-function at the origin for $V_{Q\bar{Q}}$, and
\begin{equation}
A^Q_{ZZV}=  8 \left(\frac{g_2}{c_W}\right)^2 \sqrt{3 m_V} \phi_0^Q  g^Q_A g^Q_V\  .
\end{equation}
Note that the vertex admits the Bose symmetry under  $Z(p_1,\epsilon_1)\Leftrightarrow Z(p_2,\epsilon_2)$ exchange.
By using the values\cite{Patrignani:2016xqp,Huang:2014cxa}: $M_{J/\psi}=3.0969$ GeV, $M_\Upsilon=9.4603$ GeV, $\phi_0^{J/\psi}= 0.270(20)\ ({\rm GeV})^{3/2}$, and $\phi_0^\Upsilon= 0.715(24) \ ({\rm GeV})^{3/2}$, we have $A^c_{ZZV}= 0.0866(64) \ ({\rm GeV})^2$ and  $A^b_{ZZV}= 0.723(24) \ ({\rm GeV})^2$.

Now we can calculate the SM background $Z(P_Z)\to V_Q(P_V) \nu(k_1)\bar{\nu}(k_2)$.
The differential decay width can be calculated to be
\begin{equation}
\frac{d \Gamma_{Z \to V_Q\nu\nu}}{ d x}={1\over 192\pi^3\sqrt{2}}{|A_{ZZV}|^2 G_F\over M_Z v_Q} F_{v\nu\nu}(x,v_Q)
\end{equation}
where $v_Q\equiv (M_{V_Q}/M_Z)^2$, $x\equiv (k_1+k_2)^2/M_Z^2$ and we have summed over all three neutrino species.
The function $F_{v\nu\nu}$ is given by
\begin{eqnarray}
F_{v\nu\nu}(x,v_Q)={\lambda_{cm}(x,v_Q)\over (1-x)^2(1+x-v_Q)^2}
& \times & \left[ v_Q^3(1+x)-2v_Q^2(1+6x+x^2)+\right.\nonumber \\
&& \left. v_Q(1+15x+15x^2+x^3)+4x(1-x)^2\right]\ .
\end{eqnarray}
The total branching ratios can be evaluated to be \begin{eqnarray}
Br(Z\to J/\psi \nu_i\bar{\nu_i})&=& (1.29\pm 0.10)\times 10^{-11}\,, \\
Br(Z\to \Upsilon \nu_i\bar{\nu_i})&=& (0.96\pm 0.07)\times 10^{-10} \ ,
\end{eqnarray}
 where the uncertainty arises from the wave function of quarkonium.
Again, even though the neutrinos are invisible, the invariant mass squared of two neutrinos can be reconstructed by
the energy of the quarkonium in the rest frame of $Z$:
\begin{equation}
x M_Z^2= (k_1+k_2)^2=(P_Z-P_V)^2=M_Z^2+M_V^2-2M_Z E_V^{CM}\  .
\end{equation}
This fakes the signal of $Z\to X V_Q$ production with subsequent invisible $X$ decays.

\begin{figure}
    \centering
    \includegraphics[width=0.54\textwidth]{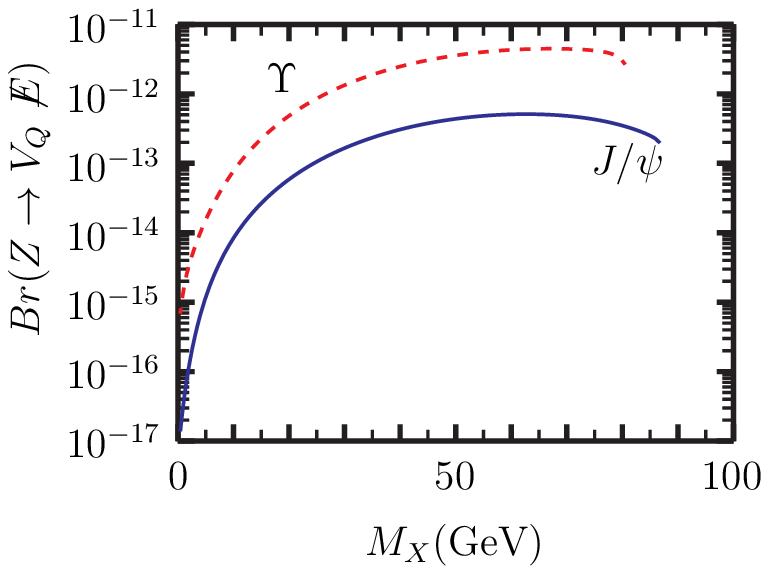}
    \caption{The SM background for $Z\ra V_Q+X$ v.s $M_X$ with a projective $1 GeV$ energy resolution at the Z-factory. Where $X$ refers to $S$ or $V_D$, and $X$ decays invisibly.
    The blue/red curve is for $Z\to (J/\psi \nu_i\bar{\nu}_i   )/(\Upsilon \nu_i\bar{\nu}_i  )$.
     }
   \label{fig:SMZ2Vnn}
\end{figure}

For $Z\to V_Q X$ with mass $m_X$, the SM background is therefore
\begin{equation}
\Delta \Gamma^{SM}_{Z\ra V_Q \nu\bar{\nu}}(m_X)= \int^{(m_X+\delta E)^2/M_Z^2}_{(m_X-\delta E)^2/M_Z^2} \left(\frac{d \Gamma_{V\nu\nu}}{dx}\right) dx
\end{equation}
 for a yet unknown detector dependent invariant mass resolution, $\delta E$.
We take $\delta E\sim 1$GeV as a conservative guess of the energy resolution at the future Z-factory, the result is displayed
in Fig.\ref{fig:SMZ2Vnn}.

 Next, we turn our attention to the SM background for the signal $Z\to V_Q X$ where $X$ decays into $\mu\bar{\mu}$. In this case one looks for the resonance peak of $m^2_{\mu\bar{\mu}}=M_X^2$.
 For $Z\ra V_Q X (X\ra\mu\bar{\mu})$, the dominant SM background comes from $Z\ra V_Q \gamma^*$  and the virtual photon turns into a muon pair.
Similar to what we did for the $ZZV_Q$, just replacing one $Z$ by a photon, the SM ``direct'' $Z-\gamma-V_Q$ coupling(amplitude) is worked out to be
\begin{eqnarray}
{\cal M}&=& A^Q_\gamma  \epsilon^{ \epsilon_Z, \epsilon_\gamma^*, \epsilon_V^*,  p_\gamma}\ , \end{eqnarray}
 with the dimensionless coupling
\begin{eqnarray}
 A^Q_\gamma &=& 8 i e_Q g^Q_A  \left(\frac{e g_2}{c_W}\right) {\sqrt{N_c M_V} \phi_0^Q \over M_Z^2(1-v_Q)}\  .
\end{eqnarray}
 Note that there is also a loop induced ``indirect'' contribution to this process \cite{Huang:2014cxa,Bodwin:2017pzj}. However its contribution is small, $\sim -5\%$ of the direct contribution and we thus ignore it in this work.

Plugging in the numbers, we have  $A_\gamma^c= 3.05(23)\times10^{-5}$ and $A_\gamma^b= 7.14(24)\times 10^{-5}$ for  $J/\Psi$ and $\Upsilon$, respectively.
The width of 2-body Z decay $Z\to f_1 f_2$ is $\Gamma = p_{cm} \langle|{\cal M}|^2\rangle/ 8\pi M_Z^2$,
where $p_{cm}=\frac{M_Z}{2} \lambda_{cm}(x_1,x_2)$ is the final state particle 3-momentum in the rest frame, and $x_i\equiv (m_i/M_Z)^2$.
Then, the total decay width of can be calculated to be:
\begin{equation}
\Gamma _{Z\ra \gamma V_{Q}}= {|A^Q_\gamma|^2 M_Z \over 96\pi  }{ (1+v_Q )(1-v_Q)^3 \over v_Q}\  .
\end{equation}
Note that these results agree with\cite{Guberina:1980dc}.
With the $Z\gamma V_Q$ vertex in hand, we can now consider the case that $X$ decays visibly, into $\mu\bar{\mu}$.
The differential width of $Z\to V_Q \mu(k_1) \bar{\mu}(k_2)$ is  straightforwardly calculated to be
\begin{equation}
\frac{d \Gamma_{Z\to V_Q \mu\mu}}{d x}={\alpha |A_\gamma^Q|^2 M_Z \over 288 \pi^2 v_Q}F_{V\mu}(x,f_\mu,v_Q)
\end{equation}
where the dimensionless variables are defined as $(k_1+k_2)^2=x M_Z^2$, $f_\mu \equiv (m_\mu/M_Z)^2$, and
\begin{eqnarray}
F_{V\mu}(x,f,v)&=& {\lambda_{cm}(x,v)\over x^2 }\{
3f^2(1+v)x \nonr \\
&&+3f[v^3-v^2(1+2x)+v(x^2+4x-1)+(1-x)^2]\nonr\\
&& + x[v^3-v^2(1+2x)+v(x^2+8x-1)+(1-x)^2] \}\ .
\end{eqnarray}

The range of $x$ for the massive fermion final state is now $  4 f_\mu\leq x\leq (1-\sqrt{v_Q})^2$.
The total branching ratios can be numerically evaluated to be
\begin{eqnarray}
Br(Z\ra J/\psi \mu\bar{\mu})&=&  (8.97 \pm 1.37)\times 10^{-10}\ , \\
Br(Z\ra \Upsilon \mu\bar{\mu})&=& (5.15 \pm 0.35)\times 10^{-10}\ .
\end{eqnarray}
Once again the uncertainties in the above arise from the wave function of quarkonium. 
Because of the photon mediation, these two branching ratios are roughly two orders of magnitude larger than the previously calculated $Br(Z\to V_Q \nu_i\bar{\nu}_i)$.
Similar to before, we integrate over the SM differential cross section at the vicinity of $x=(M_X/M_Z)^2$ with the detector energy resolution $\delta E$ which we assume to be around $\sim 1$GeV.  The SM background  for $Z\to V_Q X(X\to \mu\bar{\mu})$ is
\begin{equation}
\Delta \Gamma^{SM}_{Z\ra V_Q \mu\bar{\mu}}(m_X)= \int^{(m_X+\delta E)^2/M_Z^2}_{(m_X-\delta E)^2/M_Z^2} \left(\frac{d \Gamma_{Z\to V_Q \mu\mu}}{dx}\right) dx\ .
\end{equation}

\begin{figure}
    \centering
    \includegraphics[width=0.54\textwidth]{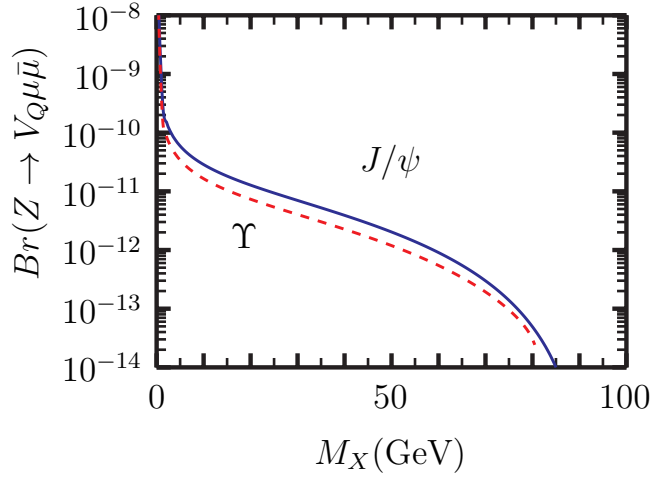}
    \caption{The SM background for $Z\to V_Q+X$ v.s $M_X$ with a projective $1 GeV$ energy resolution at the Z-factory. Where $X$ refers to $S$ or $V_D$, and $X$ decays into $\mu\bar{\mu}$.
    The blue/red curve is for $Z\to (J/\psi \mu\bar{\mu}   )/(\Upsilon \mu\bar{\mu}  )$..
     }
   \label{fig:SMZ2Vmm}
\end{figure}

Due to the photon propagator, the differential rate peaks at small $x$ (see Fig.\ref{fig:SMZ2Vmm}).
The SM background drops rapidly as $X$ gets heavier rendering the SM background basically irrelevant for $M_X>60$ GeV.

\subsection{$X =V_D$ }
\begin{figure}
    \centering
    \includegraphics[width=0.54\textwidth]{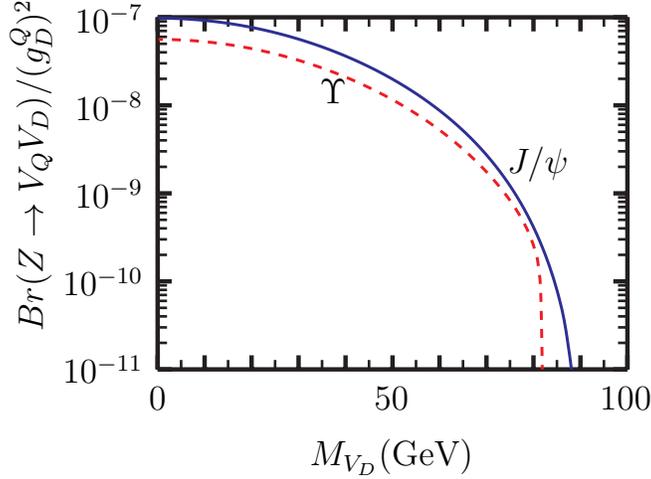}
   \caption{The Branching ratio over $g_D^2$ v.s. $M_{V_D}$. The blue/red curve is for $Z\to (J/\psi V_D )/(\Upsilon V_D )$.}
    \label{fig:ZVQVD}
\end{figure}

The $ZV_Q V_D$ coupling can be easily extended from the SM $Z V_Q\gamma$ vertex.
For the case of a light vector, we can just multiply $A_\gamma$ by $g_D^Q$ and take the light vector mass into account,
\begin{equation}
 A_{\gamma_D} = i 8 g^Q_D e_Q g^Q_A  \left(\frac{e g_2}{c_W}\right) {\sqrt{N_c M_V} \phi_0^Q \over M_Z^2(1-v_Q+d)}\ ,
\end{equation}

where $d=(M_{V_D}/M_Z)^2$.
The decay width becomes
\begin{eqnarray}
\Gamma _{Z\to V_Q V_D} &=&{(g_D^Q)^2|A_\gamma|^2 M_Z \over 96\pi v_Q }\times \left( {1-v_Q \over 1+d-v_Q}\right)^2 \times\lambda_{cm}(v_Q,d)\\
\nonumber &&\times \left\{ (1+v_Q)[(1-v_Q)^2+d^2]-2d(1-4v_Q+v_Q^2)\right\}\ ,
\end{eqnarray}
which is proportional to $(g_D^Q)^2$.
The $Z\to V_Q V_D$ decay branching ratio modulated the unknown $V_D$ coupling $(g^Q_D)^2,  Q=c,b$ is shown in Fig.\ref{fig:ZVQVD}.

\subsection{$X=S$}
\begin{figure}
    \centering
    \includegraphics[width=0.44\textwidth]{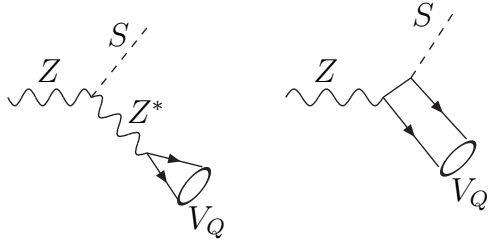}
    \caption{The Feynman diagrams for the $ZSV_Q$ coupling. Note that there are two ways of connecting the $S$ to the fermion lines. }
   \label{fig:ZSV_vertex}
\end{figure}

For $Z(P_Z,\epsilon_Z)\to V_Q(P_V,\epsilon_V) S$, there are three Feynman diagrams we need to consider, Fig.\ref{fig:ZSV_vertex}.
With the same token, the $Z S V_Q$  vertex can be calculated to be:
\begin{eqnarray}
&& \st A_s^Q M_Z \left[ -{ \epsilon_V \cdot \epsilon_Z \over 1- v_Q} + { P_V\cdot P_Z \epsilon_V\cdot\epsilon_Z - P_V\cdot \epsilon_Z P_Z\cdot \epsilon_V \over M_Z^2(1+s-v_Q) }\right]\end{eqnarray}
 with the dimensionless coupling
\begin{eqnarray}
 A_s^Q= 2 \left(\frac{g_2}{c_W}\right)^2{ \sqrt{N_c M_V} g^Q_V \phi_0^Q \over M_Z^2}\ ,
\label{eq:ZSV_vertex}
\end{eqnarray}
where $s\equiv \frac{M_S^2}{M_Z^2}$.
For $J/\Psi$ and $\Upsilon$, $A_s^c= 1.04(8)\times 10^{-5}$  and $A_s^b= -8.69(29)\times 10^{-5}$ respectively.
We have
\begin{eqnarray}
\Gamma _{Z \to V_Q S} &=&{|A_s|^2\st^2  M_Z \over 192 \pi}{\lambda_{cm}(v_Q,s) \over v_Q}
\left\{ {(1+v_Q-s)^2+8v_Q\over (1-v_Q)^2}\right.  \nonr \\
&&\left. -{12v_Q(1+v_Q-s)\over(1-v_Q)(1+s-v_Q)}+{2v_Q[(1+v_Q-s)^2+2v_Q]\over(1+s-v_Q)^2}\right\}\ .
\label{eq:ZSV_rate}
\end{eqnarray}
The first term in the curvy bracket represents the contribution of the diagram where $S$ connects to the Z boson.
The third term represents the contributions where $S$ connects to the quark lines in the quarkonium.
And the middle term is the interference contribution.
Note that, $\st$ aside, there is a sign difference for the first $ \epsilon_V \cdot \epsilon_Z$ term in Eq.(\ref{eq:ZSV_vertex})
between us and \cite{Guberina:1980dc} due to a difference in convention.
However, for the decay width, Eq.(\ref{eq:ZSV_rate}) agrees with \cite{Guberina:1980dc}.

\begin{figure}
    \centering
    \includegraphics[width=0.54\textwidth]{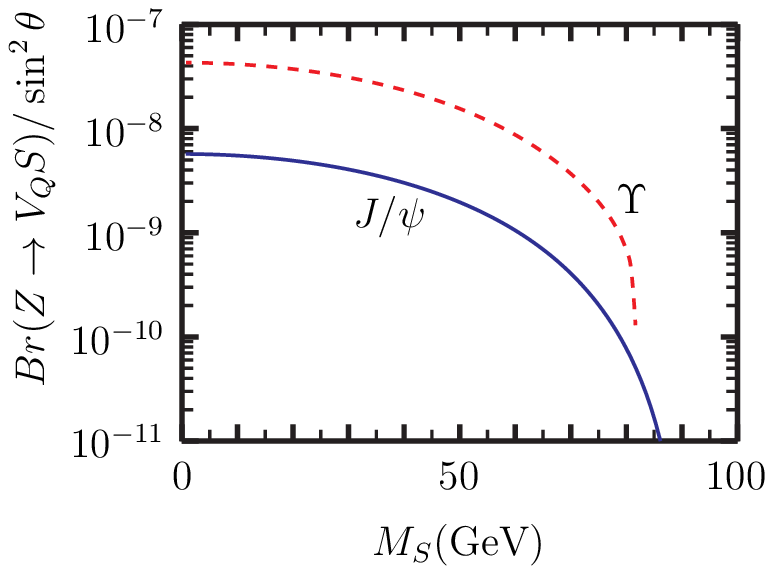}
    \caption{The Branching ration over $\st^2$ v.s. $M_S$. The blue/red curve is for $Z\to (J/\psi S )/(\Upsilon S )$.}
   \label{fig:Z2VQS}
\end{figure}

The $Br(Z\to S V_Q)/\st^2$ is displayed in Fig.\ref{fig:Z2VQS}.
Note that the $Br(Z\to \Upsilon S)$ is about one order of magnitude larger than $Br(Z\to J/\psi S)$. This can be understood due to
the ratio $(g_V^b/g_V^c)^2\cdot (M_\Upsilon/M_{J/\psi})\cdot(m_b/m_c)^3\sim 10$.

\section{When $S$ does not mix}\label{sec:vevless}
Let us briefly consider the case where $S$ does not develop a vacuum expectation value(VEV). In this case the only coupling to the standard model is through the interaction with the Higgs.
After spontaneous symmetry breaking we have a dimensionful coupling between the singlet and the Higgs which controls the decay rate. The relevant interaction is parameterized as
\begin{equation}
\frac{\kappa v_H}{2} H S^2
\end{equation}
where $v_H=246$ GeV is the  VEV of the SM Higgs.
Then the $H\to 2S$ decay width can be calculated to be
\begin{equation}
\Gamma_{H\ra 2S}= {\kappa^2 v_H^2 \over 32 \pi M_H}\sqrt{1-(4M_S^2/M_H^2)}
\end{equation}

 The above contributes to the Higgs invisible decay.  ATLAS\cite{Aad:2015txa} and CMS\cite{CMS:2015naa} give limits on the Higgs invisible decay of $Br_{inv}<0.28$ and $Br_{inv}<0.36$ at 95\%CL respectively.
Using the ATLAS bound, it amounts to $\Gamma_{inv}<0.388 \Gamma_{visibale}= 1.58 MeV$ for the SM 125GeV Higgs.
This translates to an upper bound on the triple scalar coupling of
\begin{equation}
\kappa^2 <{1\over\sqrt{1-4M_S^2/M_H^2}}{32 \pi M_H \Gamma_{inv} \over v_H^2}.
\end{equation}
The coupling is severely constrained to be smaller than  $\lesssim 4\times 10^{-4}$ for the mass range we are interested in, $M_S< M_H/2$.
The widths of $Z\to \mu\bar{\mu} S S$ can be calculated by CalcHep.
For $M_S= 5(20)$GeV, the width is $2.3\times 10^{-11}(2.01\times 10^{-12}) \times (\kappa^2/10^{-4}) $ GeV.
And the SM background will be $Z\to \mu\bar{\mu} \nu\bar{\nu}$ and the width is $2.662\times 10^{-8}$ GeV.
Given $10^{12}$ Z bosons, the number of expected signal events is around ${\cal O}(1)$ and the expected number background events is around $10^4$.
Hence, the Z factory can not compete with the Higgs invisible decay constraint in this scenario.

Similar excise can be carried out for the Higgs factory. We find the signal to background ratio is smaller than $10^{-6}$.
Moreover, even with a luminosity around $ab^{-1}$, the expected signal event number is $\lesssim {\cal O}(1)$.

\section{Results and Conclusions}\label{sec:conclusions}
 \begin{figure*}[htb!]
 \centering
 \subfigure[]{\includegraphics[width=0.44\textwidth]{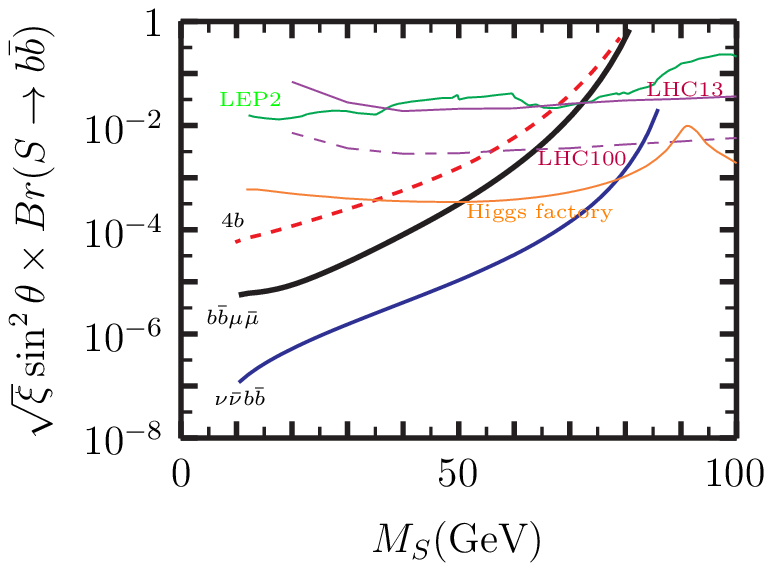}} \quad
  \subfigure[]{\includegraphics[width=0.44\textwidth]{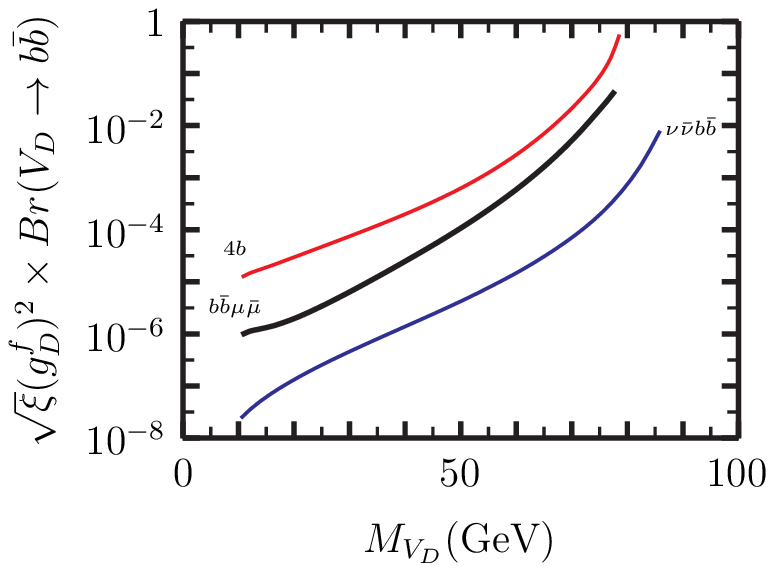}} \\
   \subfigure[]{\includegraphics[width=0.44\textwidth]{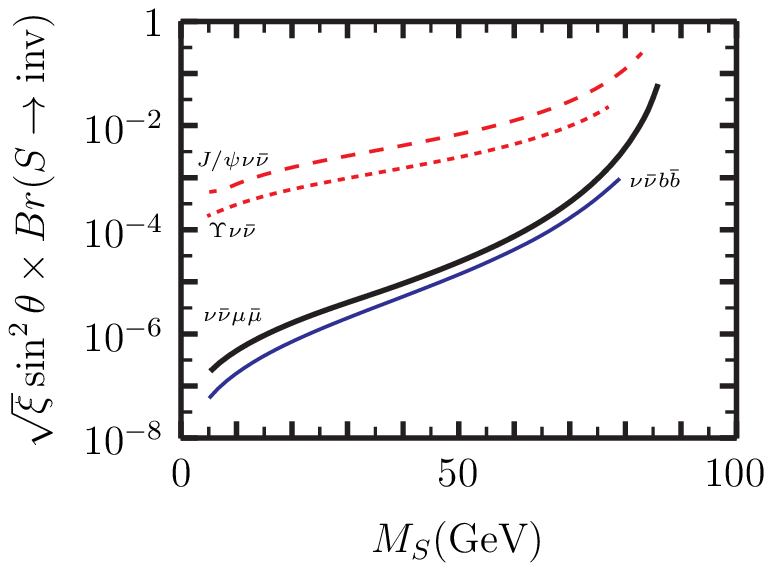}} \quad
 \subfigure[]{\includegraphics[width=0.44\textwidth]{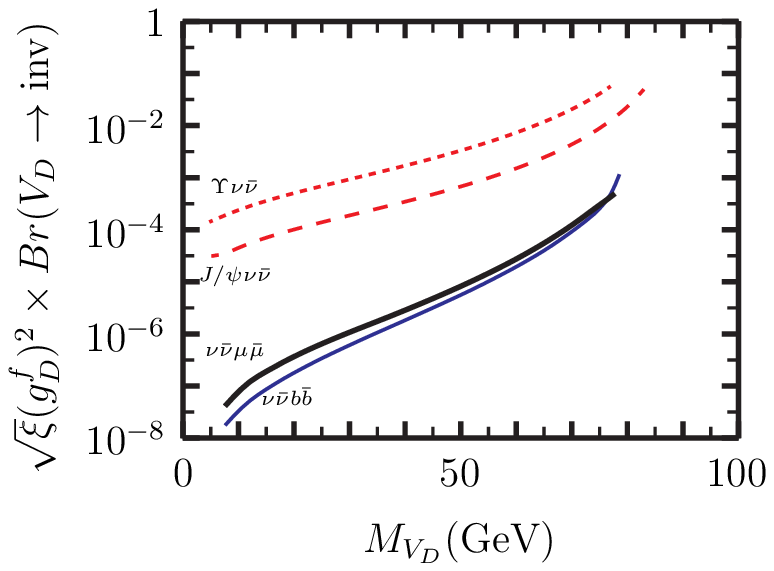}} \\
 \subfigure[]{\includegraphics[width=0.44\textwidth]{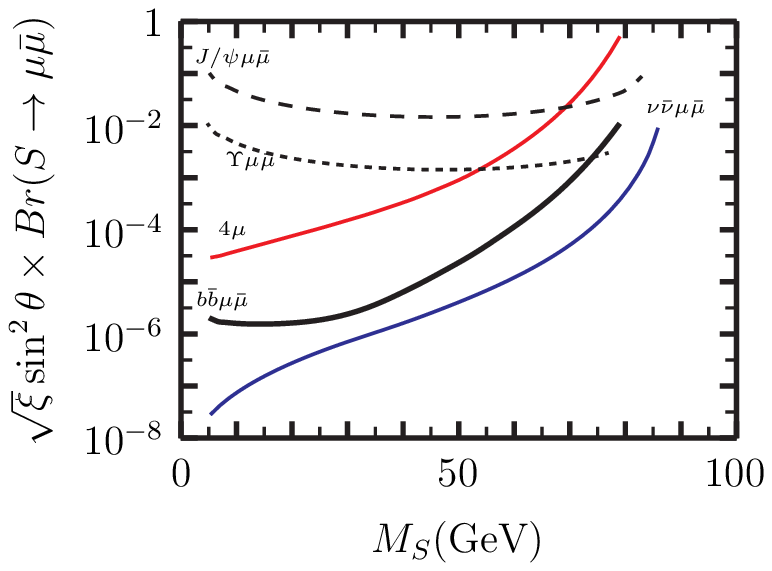}}\quad
  \subfigure[]{\includegraphics[width=0.44\textwidth]{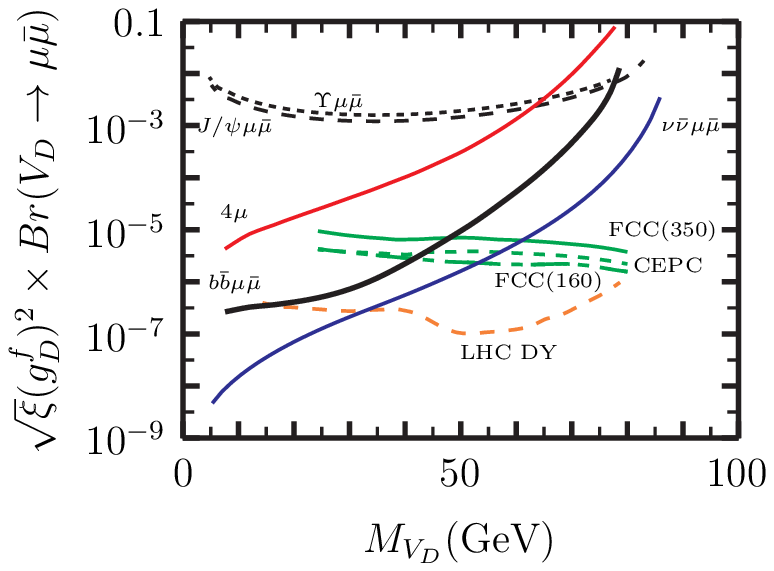}}
\caption{  (a,c,e) The $3 \sigma$  limits on $\sqrt{\xi}\st^2\times Br(S\to f\bar{f})$ v.s. $M_S$ with $10^{12}$ $Z$'s. (b,d,f) The $3 \sigma$  limits on $\sqrt{\xi}(g_D^f)^2\times Br(V_D\to f'\bar{f'})$ v.s. $M_{V_D}$ with $10^{12}$ $Z$'s and energy resolution $\triangle E=1$GeV.
Here $\xi$ denotes the unknown total detection efficiencies for the signals.
In (a), we also display the limits from the direct search at LEP2\cite{Barate:2003sz}(where the detection efficiencies have been taken into account), $pp\ra t\bar{t}S(b\bar{b})$ at HLC13 and HLC100 with ${\cal L}=3 ab^{-1}$, and at the Higgs factory\cite{Chang:2017ynj} with ${\cal L}=1 ab^{-1}$. In (f), some limits adopted from the Drell-Yan process at LHC\cite{Curtin:2014cca} and $e^+e^-\ra \gamma \mu\bar{\mu}$ at the CEPC and FCC\cite{He:2017zzr} are displayed. Note however these limits apply to the kinetic $U(1)_Y\mhyphen U(1)_{\mbox{hidden}}$ mixing model only.   }
\label{fig:limit_Xff}
 \end{figure*}
\begin{figure*}[htb!]
 \centering
 \subfigure[]{\includegraphics[width=0.44\textwidth]{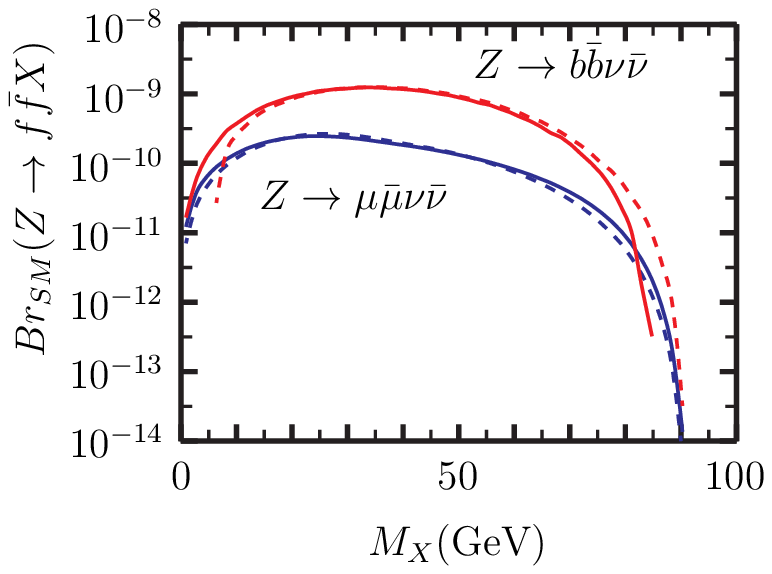}} \quad
 \subfigure[]{\includegraphics[width=0.44\textwidth]{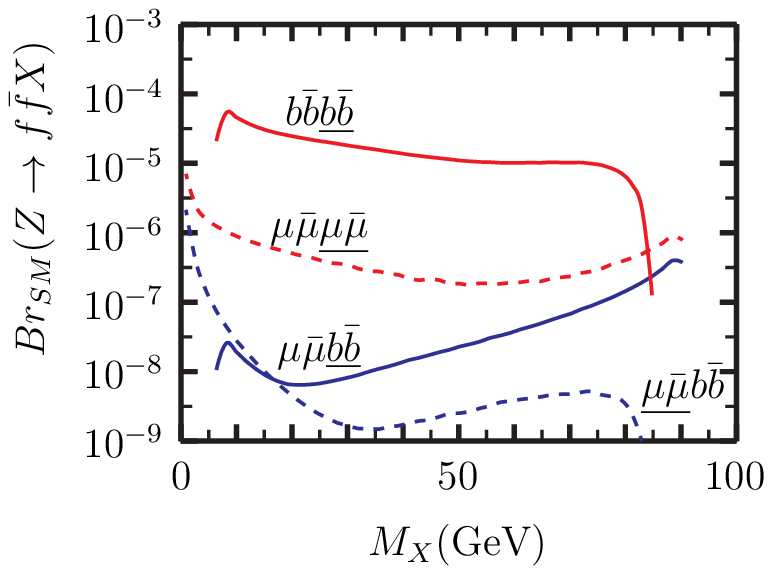}}\quad
\caption{(a) The SM background decay branching  vs $M_X$.
    The solid/dash line is for the $X$ invisible/visible decay associated with $f=\mu, b$.
    (b) The SM background decay branching for visible decay associated with $\mu$ or $b$ vs $M_X$. }
\label{fig:SMBG_Z4f}
 \end{figure*}

In this work we have studied the possibility of probing the parameter space for light boson, $X$, extensions to the standard model with a Z factory. We have focused on the rare Z decays  $Z \to f \bar{f} X$ and $Z \to V_Q X$ where the fermions are either invisible states, b-quarks or muons. These states are useful probes due to  the increasing efficiency of b tagging and relatively long life time respectively.
Other channels are less useful for such a search. In particular, the light jets have a noisy background, and the $\tau$ lepton has multiple hadronic decay channels rendering it more difficult to reconstruct. Moreover, our formulas  can be easily extended to these cases.
The SM 4-body $Z$ decay backgrounds listed in Tab.\ref{tab:SMBG} are evaluated by CalcHep and displayed in Fig.\ref{fig:SMBG_Z4f}.

 For the $Z\to V_Q X$ background $ Z\to V_Q f\bar{f}$, with $f=\mu,\nu$, we derived  the analytic expressions and the numerical evaluation of these expressions are displayed in Fig.\ref{fig:SMZ2Vnn} and Fig.\ref{fig:SMZ2Vmm}.  To avoid complications arising from QCD, we do not consider the channel $Z\to V_Q b\bar{b}$.

We now have all the ingredients to compare signals with SM backgrounds.
In Fig.\ref{fig:limit_Xff}, we plot the curves corresponding to the $3\sigma$ limit that satisfies
\begin{equation}
\frac{S}{\sqrt{B}}= {Br(\mbox{signal}) \times 10^6 \times  \sqrt{\xi}\over \sqrt{Br(\mbox{background},\triangle E=1\mbox{GeV})\times(\triangle E /1\mbox{GeV} )}}=3\ ,
\end{equation}
with $10^{12}$ fiducial $Z$ events and  final state invariant mass resolution $\triangle E=1$ GeV\footnote{The expected precision of invariant mass  at the planning Z-factories (see Fig.3.15 and Fig.3.16 in\cite{CEPC-SPPCStudyGroup:2015csa}) are  about $\delta m_{b\bar{b}}\sim 2$GeV and $\delta m_{\mu\bar{\mu}}\sim 0.2$ GeV for the $b\bar{b}$ and $\mu\bar{\mu}$, respectively.  }.
  We do not include the uncertainty in the quarkonium wave function, $\sim {\cal O}(10\%)$, for the processes involving $V_Q$, nor the
non-resonant SM backgrounds, $\sim {\cal O}(10\%)$, as their effects are barely visible in the figure.
Here we also included the unknown overall detection efficiency $\xi$ for a specific channel. 
For those channels with SM background branching ratios smaller then $10^{-12}$, we use $10^{-12}$ as the background.

When the energy resolution is reasonably small, the number of background events is linearly proportional to the energy resolution.
For a different energy resolution,  the corresponding $3\sigma$ limits can be easily  read from  Fig.\ref{fig:limit_Xff} with a vertical shift $\frac{1}{2} \log_{10}(\triangle E/ 1\mbox{GeV})$.

For quarkonium, $V_Q\ra l\bar{l}$ offers a clean tag for the particle identity and the di-lepton decay branching ratios, $Br(J/\Psi\ra l\bar{l})=11.932(46)\%$ and $Br(\Upsilon\ra l\bar{l})=7.46(15)\%$\cite{Patrignani:2016xqp} are well measured.
Therefore, the dominant factor of the quarkonium identification is the detection efficiencies of di-lepton, and the overall $\xi$ will be similar to
that of the four charged leptons final state.
The expected $e(\mu)$ identification efficiency at the CEPC is $99.5\%(98.5\%)$ for charged lepton energy $E>10$ GeV, and it
drops to $\sim 96\%(85\%)$ when $E\sim 2$GeV\cite{CEPC-SPPCStudyGroup:2015csa}. From these numbers, a simple estimation is that $\xi\sim (0.985)^4=0.94$ when all four muon energies are larger than $10$GeV. For the events with one muon energy around $2$GeV and the other three $>10$GeV, $\xi\sim(0.85)\times(0.985)^3=0.81$.
For b quark, the tagging efficiency is $\sim 90\%$ \cite{CEPC-SPPCStudyGroup:2015csa}\footnote{ At LEP, the efficiencies for b-tagging range from $21.0\%$ to $44.0\%$ depending on the b-purities\cite{Abdallah:2002xm}. With the implementation of new techniques like neural networks and boosted decision trees, an efficiency to identify $b$ jet of $70\%$ can be achieved at the ATLAS\cite{Aad:2015ydr}.  }.
Thus, $\xi\sim (0.9)^4= 0.66$ for the $4b$ channel. For the $\mu\bar{\mu} b\bar{b}$ channel, the overall detection efficiency
ranges roughly from $\xi\sim 0.68$, with one low energy muon, to $0.78$ when all particle energies are larger than $10$ GeV.
Of course, to determine the actual detection efficiency for a specific channel, a comprehensive analysis of the full kinematic and the detector performance is needed and we should leave it to the future studies.

 As have been studied in \cite{Chang:2016pya,Chang:2017ynj}, the $Z\to S(b\bar{b}) f\bar{f}$ is a very promising channel for either discovery or falsifying a sizeable portion of the light scalar parameter space, Fig.\ref{fig:limit_Xff}(a).
This new  $\st^2\times Br(S\ra b\bar{b})$ limit  out performs the LEP-II limits\cite{Barate:2003sz} ( the green curve in Fig.\ref{fig:limit_Xff}(a) )  by around five/one orders of magnitude
at $M_S=10/80$ GeV by using the $Z\ra \nu\bar{\nu}b\bar{b}$ channel due to the enormous number of $Z$s that are expected to be produced.
The signal will be a sharp b-pair invariant mass in the decay $Z\to \nu\bar{\nu} b\bar{b}$.
This channel is useful for probing $S$ with large $b\bar{b}$ ( or $\mu\bar{\mu}$ when $M_S<2m_b$, Fig.\ref{fig:limit_Xff}(e)) decay fraction.
A similar limit is obtained for the light vector, Fig.\ref{fig:limit_Xff}(b).
In the Higgs portal models, the Yukawa couplings of the singlet scalar are proportional to the fermion masses.
Thus it is expected that the constraint on the mixing $\st^2$ from the visible decay mode $S\ra \mu \bar{\mu}$ is four orders of magnitude weaker than that of the $S\ra b \bar{b}$ mode if the detector has similar energy resolutions on determining the invariant masses $m_{b\bar{b}}$ and $m_{\mu\bar{\mu}}$.
On the other hand, the new vector boson couplings in general are not proportional to the fermion masses.
Therefore, whether $V_D\ra b\bar{b}$ or $V_D\ra \mu\bar{\mu}$ are better visible channels for discovery will
depend on the unknown detector performance in the future machine.
The searching strategies we proposed to look for the vector boson with mass $1\lesssim M_{V_D} \lesssim 80$ GeV at the Z factories is model independent. The $Z\to f\bar{f} V_D$
signals are determined by the phenomenological flavor dependent
gauge coupling $f_D^f$, assuming there is no mixing between the SM Z boson and the physical state $V_D$. On the other hand, the limits obtained in \cite{Hoenig:2014dsa, Curtin:2014cca}, where the kinematic mixing parameter can be probed down to $\epsilon\sim 10^{-3}$ by electroweak precision or the Drell-Yan processes at the LHC, requiring a UV complete model. For instance, the Drell-Yan processes need the vector boson couplings both to quarks and leptons at the same time.

 Moreover, we found  that the invisible decay of $X$  can be utilized at the Z-factory with precision controlled $e^\pm$ beam energy, Figs.\ref{fig:limit_Xff}(c) and Fig.\ref{fig:limit_Xff}(d).
This mode provides a powerful handle to probe the $X$ with sizable invisible decay branching ratios, for example the model discussed in \cite{Chang:2016pya}.
 We know of no other reactions that can compete with this for the mass range of $X$ we are studying.
Furthermore, we have also studied the 2-body $Z\to \gamma X$ process with the production of back-to-back  on-shell $X$ and a high energy photon and found that the signal cannot compete with the SM background, $e^+e^-\to \gamma f\bar{f}$.  However, with a higher center mass energy, $\sqrt{s}\gtrsim 160$GeV, this $e^+e^-\to \gamma f\bar{f}$ channel is useful for detecting a vector boson with mass $\gtrsim 20$GeV  which kinematically mixes with the SM $U(1)_Y$ \cite{He:2017zzr}.  In this model, the $Br(V_D\ra f\bar{f})$ is determined by the strength of kinetic mixing parameter $\epsilon$.
When $M_X< v_H$, the $X$-fermion coupling is mostly vector-like and therefore $g^f_D\simeq \epsilon$.
 Their $2\sigma$ limits on $\epsilon^2$ multiplied by $Br(V_D\ra \mu\bar{\mu})$ for $f=\mu$ with ${\cal L}=1 ab^{-1}$ at the FCC-ee(160GeV,350GeV) and CEPC(240GeV) are shown in Fig.\ref{fig:limit_Xff}(f). A different limit from the Drell-Yan process $pp\ra V_D\ra ll$ at LHC14 with ${\cal L}=3 ab^{-1}$\cite{Curtin:2014cca}
is displayed alongside.
 We stress that these limits apply solely to the specific $U(1)$ kinetic mixing model.

 It is also worth pointing out that, since there is no meaningful SM background, the above mentioned searches can set stringent constraint on the vector boson flavor-changing decay branching ratios $Br(V_D\to f_i\bar{f_j} +f_j \bar{f_i})$ as well. For example, by using the signal $Z\to \mu\bar{\mu} V_D(V_D\to f_i\bar{f_j} +f_j \bar{f_i})$, the combination $(g_D^\mu)^2\times Br(V_D\to f_i\bar{f_j} +f_j \bar{f_i})$ can be probed to the $10^{-9}(10^{-3})$ level for $M_{V_D}\sim$ a few ($80$) GeV.
Additionally, in the case of an inert singlet scalar, which does not mix with the SM Higgs, the Z and Higgs factory can not compete with the limit from Higgs invisible decay \cite{Aad:2015txa,CMS:2015naa}.
We also considered the possibility of probing a light $X$ boson by utilizing the $Z\ra V_Q X$ channel, where $V_Q= J/\Psi, \Upsilon$.
 Although the limits are relatively weak, this process provides additional search strategy for $X$ with  a large invisible ( or $\mu\bar{\mu}$ ) decay branching ratio  and a cross check if $X$ is found, Figs.\ref{fig:limit_Xff}(c-f).
 On the other hand, the light exotic vector possibility will be excluded if the signal $Z\ra \Upsilon \mu\bar{\mu}$ with a $\mu\bar{\mu}$ resonance is seen but the counterpart signal $Z\ra b\bar{b} \mu\bar{\mu}$ is not.

Finally, we note that we can probe regions of parameter space with significantly smaller branching ratios to SM particles than has previously been considered. As such it is worth the caution to check to see if the life time of the singlet scalar can ever be so long as to fake an invisible decay. We find that for our range of masses the mixing angle needs to be several orders of magnitude below what we consider before the singlet life time becomes an issue.

Of course, we cannot predict the details and parameters of the future machines. Our numerical estimations in this work should be regarded as explorative speculation at this moment. The realistic analysis is still awaited for the experimentalist to perform in the future.
\begin{acknowledgments}
 WFC is supported by the Taiwan Minister of Science and Technology under
Grant Nos. 106-2112-M-007-009-MY3  and 105-2112-M-007-029.
TRIUMF receives federal funding via a contribution agreement with the National Research Council of Canada and the Natural Science and Engineering Research
Council of Canada.
\end{acknowledgments}

\bibliographystyle{JHEP}
\bibliography{references}

\end{document}